\documentclass[journal=jacsat,manuscript=article]{achemso}
\usepackage[version=3]{mhchem}
\usepackage{graphicx}
\usepackage{bm}
\usepackage{xcolor}
\usepackage{booktabs}
\usepackage{tabu}
\usepackage{float}
\usepackage{braket}
\usepackage{multirow}
\usepackage{xr}
\makeatletter
\newcommand*{\addFileDependency}[1]{
  \typeout{(#1)}
  \@addtofilelist{#1}
  \IfFileExists{#1}{}{\typeout{No file #1.}}
}
\makeatother

\newcommand*{\myexternaldocument}[1]{
    \externaldocument{#1}
    \addFileDependency{#1.aux}
}

\myexternaldocument{SI}
\author{Guorong Weng}
\author{Mariya Romanova}
\author{Arsineh Apelian}
\author{Hanbin Song}
\author{Vojt\v{e}ch Vl\v{c}ek}%
\email{vlcek@ucsb.edu}
\affiliation{Department of Chemistry and Biochemistry, University of California, Santa Barbara, CA 93106-9510, U.S.A.}%

\title{Reduced scaling of optimal regional orbital localization via sequential exhaustion of the single-particle space}


\begin{document}

\begin{tocentry}
    \centering
    \includegraphics[width=4.5cm]{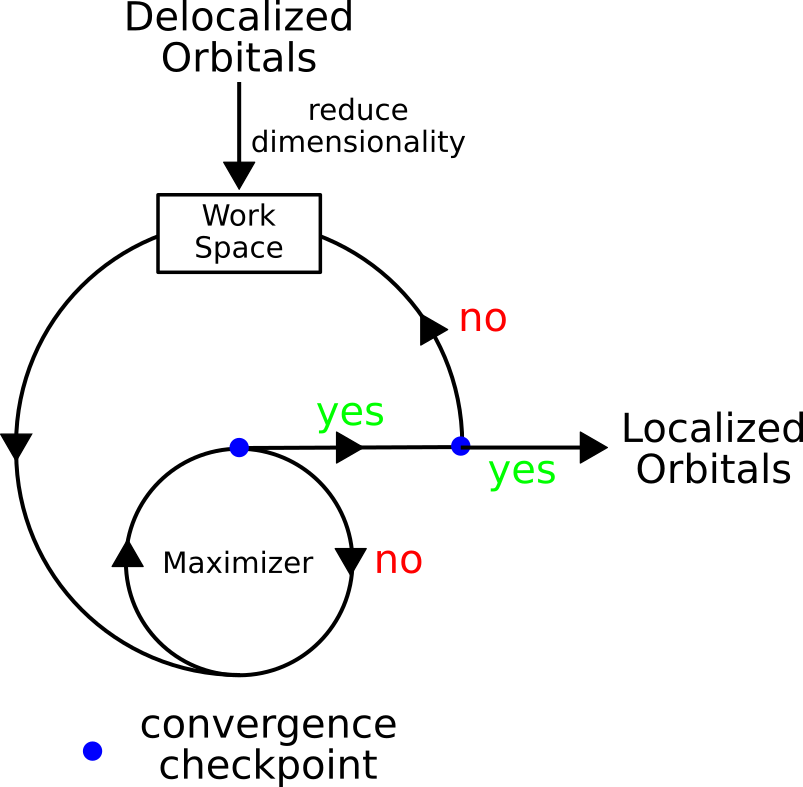}
    \label{fig:TOCentry}
\end{tocentry}

\begin{abstract}
Wannier functions have become a powerful tool in the electronic structure calculations of extended systems. The generalized Pipek-Mezey Wannier functions exhibit appealing characteristics (e.g., reaching an optimal localization and the separation of the $\sigma$-$\pi$ orbitals) when compared with other schemes. However, when applied to giant nanoscale systems, the orbital localization suffers from a large computational cost overhead when one is interested in localized states in a small fragment of the system. Herein we present a swift, efficient, and robust approach for obtaining regionally localized orbitals of a subsystem within the generalized Pipek-Mezey scheme. The proposed algorithm introduces a reduced workspace and sequentially exhausts the entire orbital space until the convergence of the localization functional. It tackles systems with $\sim$10000 electrons within 0.5 hours with no loss in localization quality compared to the traditional approach. Regionally localized orbitals with a higher extent of localization are obtained via judiciously extending the subsystem's size. Exemplifying on large bulk and a 4-nm wide slab of diamond with NV$^-$ center, we demonstrate the methodology and discuss how the choice of the localization region affects the excitation energy of the defect. Furthermore, we show how the sequential algorithm is easily extended to stochastic methodologies that do not provide individual single-particle eigenstates.  It is thus a promising tool to obtain regionally localized states for solving the electronic structure problems of a subsystem embedded in giant condensed systems.
\end{abstract}

\maketitle

\section{\label{sec:intro}Introduction}

Localized orbitals are widely used in electronic structure computations for multiple purposes: conceptually, they can provide valuable information about chemical bonding and chemical properties of molecules and materials. More importantly, they allow the evaluation of non-local two-body interaction integrals at a significantly reduced cost due to the reduced spatial overlaps. Hence, they represent a powerful tool in mean-field and post-mean-field electronic structure calculations such as hybrid functional calculations\cite{Wu2009,Gygi2013}, density functional theory with the Hubbard correction term\cite{Miyake2009,Tomczak2009}, or many-body caluations.\cite{Choi2012,Weng2021} In the same vein, the maximally localized orbital descriptions is optimal for treating correlation phenomena since (due to the locality) the number of ``inter-site'' interactions is minimal, and the effective size of the problem is smaller. As a result, optimally localized states are essential in the context of embedding and downfolding for many-electron problems\cite{Aryasetiawan2009,Pvarini2011,Bowler2012,Lau2021}. 

Arguably, the most popular approaches to obtain localized states are the Foster-Boys (FB) scheme \cite{Boys1960,Foster1960} for molecules and the maximally-localized Wannier functions (MLWF)\cite{Marzari1997,Marzari2012} for periodic solids. In addition, selected columns of the density matrix (SCDM)\cite{Damle2015} localization scheme provides an alternative to find a localized basis,\cite{Damle2017,Damle2018,Damle2019,Vitale2020} which however avoids the localization optimization. Among these and other orbital localization schemes\cite{Edmiston1963,Edmiston1965,Lowdin1966,Niessen1972,Damle2015}, the Pipek-Mezey (PM) localized molecular orbitals \cite{Pipek1989} are appealing due to their high spatial  localization and (conceptually) the separation of $\sigma-\pi$ characters of chemical bonds. Recently, PM localized molecular orbital formalism has been expanded to periodic systems\cite{Jonsson2017}. This generalized Pipek-Mezey Wannier Functions (G-PMWF) approach retains the advantages (particularly stronger localization) compared with MLWF. 

The iterative optimization, however, translates to a high computational cost \cite{Pipek1989} and requires that all single-particle states are known. This becomes a bottleneck for giant systems: the overhead is substantial when one is interested only in a small portion of the otherwise giant system, such as maximally localized orbitals associated with a point defect in solids, an adsorbate molecule on a surface, or molecular states in a complex environment. Here, it is still often necessary to handle the entire problem, despite only a fraction of localized states being sought. Such nanoscale problems involve thousands of electrons and the prevalent strategy is to lower the number of iteration steps necessary to reach the optimum, e.g., by a robust solver\cite{Clement2021}. Although the proposed scheme can effectively lower the iteration steps towards convergence, a minimal atomic orbital basis is still required for determining the atomic charges. Further, for truly giant systems, one would employ techniques that avoid the use (or knowledge) of all single-particle states\cite{Baer2013,Cytter2018,Chen2019,Fabian2019,Nguyen2021,Baer2022,Neuhauser2013_RPA,romanova2022stochastic,Neuhauser2013_MP2,Ge2014,Neuhauser2017,Dou2019,Takeshita2019,neuhauser2014breaking,vlcek2017stochastic,Vlcek2018swift,romanova2020decomposition,vlcek2019stochastic}. 

Herein, we present a new and complementary top-down approach leading to a fast, efficient, and robust orbital localization algorithm via sequentially exhausting the entire orbital space. It is beneficial for obtaining regionally localized orbitals for a subsystem within the G-PMWF scheme. In contrast to other methods, the problem's dimensionality is reduced from the outset by partitioning the orbital space. As our work space is effectively compressed, the dimensionality of the relevant matrices in the G-PMWF scheme is much smaller and therefore, the time per iteration step is shortened by orders of magnitude. The unitary transform is performed iteratively till convergence. The transformation starts directly either with: (i) the canonical delocalized orbitals without any external or auxiliary atomic basis set\cite{Zhang2014,Clement2021}; or (ii) initial guess of the subspace of localized single-particle orbitals (which can be obtained by, e.g., by filtering\cite{Baer2013,Chen2019,Fabian2019,Baer2022,neuhauser2014breaking,Vlcek2018swift}). The compression of dimensionality helps to reduce the scaling of the method with the number of electrons to be linear. The completeness of sequentially exhausting the orbital space is demonstrated by the converged localization functional. We further test the quality of the localized basis by constructing an effective Hubbard model for the NV$^-$ defect center in diamond and computing its optical transition energies in bulk supercells and a giant (4~nm thick) slab containing nearly 10000 electrons. Excellent agreement between the sequential exhausting approach and the full space approach is achieved for the computation of optical transition energies. The accuracy of Hubbard model calculations is further improved by the Wannier function basis obtained from the subsystem with an extended size. In the last section, we provide a thorough discussion on how the choice of localization affects the excitation energies of the embedded NV$^-$ center.

\section{\label{sec:theory}Theory}

\subsection{Generalized Pipek-Mezey Wannier Functions}
In this subsection, we briefly revisit the G-PMWF formalism\cite{Jonsson2017} to clarify the motivation for this work.
The G-PMWF evaluates the orbital localization using the following quantity $\mathcal{P}$ as a functional of the unitary matrix $\textbf{U}$:
\begin{equation}\label{eq:p}
\mathcal{P(\textbf{U})} = \sum_{i=1}^{N_{s}} \sum_{A=1}^{N_A} [Q_{ii}^A(\textbf{U})]^2.
\end{equation}
Here, $i$ denotes the $i^{\rm th}$ state, and $N_{s}$ represents the number of states that spans a particular orbital space. $A$ denotes the $A^{\rm th}$ atom in the system, and $N_A$ is the number of atoms in the system. $Q$ is termed the atomic partial charge matrix (defined below). In practice, $Q_{ii}^A$ represents the partial charge on atom $A$ contributed by state $i$. The stationary point of $\mathcal{P}$ corresponds to the unitary matrix $\textbf{U}$ that transforms the canonical states into Pipek-Mezey localized states
\begin{equation}\label{eqn:pmwf}
\ket{\psi^{PM}_j} = \sum_k^{N_{s}} \textbf{U}_{jk} \ket{\phi_k},
\end{equation}
where $\ket{\phi_k}$ represents the canonical state.

Generally, the value of $\mathcal{P}$ is iteratively maximized till reaching convergence. In the $n^{\rm th}$ iteration step, the $Q$ matrix is defined as
\begin{equation}\label{eq:apcm}
Q^{A,n}_{ij} = \braket{\psi_i^{n}|w_A|\psi_j^{n}}.
\end{equation}
Here $\ket{\psi_i^n}$ represents either the transformed state ($n>$0) or the canonical state ($n$=0). In real-space representation, $w_A$ denotes the atomic weight function\cite{Hirshfeld1977,Cioslowski1991,bader1994,Alcoba2006,Timothy2008,Timothy2009,Knizia2013,Janowski2014,Lehtola2014,Knizia2015} in replacement of the Mulliken partial charge scheme\cite{Pipek1989}.

For $n\ge1$, the $Q$ matrix can also be calculated by
\begin{equation}\label{eq:apcm2}
Q^{A,n}_{ij} = \sum_k^{N_{s}} \sum_l^{N_{s}} ({\textbf{U}^{n}}^{\dagger})_{ik} Q^{A,n-1}_{kl} \textbf{U}^{n}_{lj}.
\end{equation}

Note that in practice, the $Q$ matrix has a dimensionality of $N_A\times N_{s}^2$. The number of elements can rocket to 10$^9$ for a system with 10$^3$ atoms with $10^3$ occupied states. Furthermore, the theoretical scaling of the method is $N_{g}\times N_{s}^2$ ($N_{g}$ denotes the number of grid points in real-space). Our numerical results for the defect center in diamond are close to this theoretical behavior, as discussed in the Results and Discussion section.

\subsection{Sequential Variant of G-PMWF}

This subsection presents an efficient algorithm to obtain a subset of PMWFs localized on a specific set of atoms.

\subsubsection{Fragmentation treatment}
Conventionally, one has to localize all $N_{s}$ states and then identify $N_{rl}$ states that are regionally localized on the selected atoms. For instance, for a $\text{CH}_4$ molecule surrounded by other atoms/molecules, $N_{rl}$ will be four if considering only the valence electrons and doubly occupancy.
When $N_{rl}\ll N_{s}$, this approach suffers from a significant overhead. This is quite limiting when nanoscale systems are considered: the dimensionality of matrix $Q$ and the computational scaling make it challenging to work with thousands of electrons. Previously, we introduced a modified form of the PM functional to account for $N_A'$ ($N_A'\ll N_A$) selected atoms only and search for the $N_{rl}$ states directly\cite{Weng2021} . Such a modification is equivalent to the search of a local maximum of $\mathcal{P}$ on the selected atoms, and it reduces the dimensionality to $N_A'\times N_{s}^2$. In this work, we further compress the $N_A'$ to simply 1 by creating a single fragment from the subset of atoms. Unlike the ``fragment" proposed in the FB scheme\cite{Zhang2014}, our definition of a fragment is defined using the atomic weight function $w_A$
\begin{equation}\label{eq:fwf}
w_f(\textbf{r}) = \sum_A^{N_A'} w_A(\textbf{r}),
\end{equation}
where $f$ denotes the fragment of interest. The functional $\mathcal{P}$ thus becomes
\begin{equation}\label{eq:Pprime}
\mathcal{P'(\textbf{U})} = \sum_i^{N_{rl}}[Q_{ii}^f(\textbf{U})]^2,
\end{equation}
where $\mathcal{P}'$ is the modified PM functional for the fragment. 

Note that: (i) the unitary transform is still performed on \textit{all} $N_{s}$ states that need to be known, and (ii) the $N_{rl}$ states are identified from $N_s$ by evaluating the partial charge on the selected fragment. In this context, we define the measure of the locality of a specific state on the fragment as
\begin{equation}\label{eq:localityfrag}
L^f_i = \braket{\psi_i|w_f|\psi_i}.
\end{equation}
Its value ranges from 0 (not localized) to 1 (most localized). Only the top $N_{rl}$ states of the $N_{s}$ states in the decreasing order of $L^f_i$ are considered the regionally localized Wannier functions on the fragment.

Next, the G-PMWF approach is broken into two steps: (1) maximize $\mathcal{P}'$ (Eq.\eqref{eq:Pprime}) and find the $N_{rl}$ states that are localized on the fragment; (2) maximize the canonical $\mathcal{P}$ (Eq.\eqref{eq:p_2}) using the $N_{rl}$ states from step 1 and obtain localized states on each individual atom of the fragment. 
\begin{equation}\label{eq:p_2}
\mathcal{P(\textbf{U})} = \sum_{i=1}^{N_{rl}} \sum_{A=1}^{N_A'} [Q_{ii}^A(\textbf{U})]^2.
\end{equation}
Essentially, the first step is a ``folding" step where the electron density is effectively localized on the fragment disregarding the individual atoms. The second step is instead an ``unfolding" step where the electronic states obtained from step 1 are unfolded onto each individual atom in the fragment.

The $Q$ matrix is reduced to $N_{s}^2$ in step 1 and to $N_A' \times N_{rl}^2$ in step 2, respectively. The second step is trivial in cost since $N_{rl}$ is often much smaller than $N_{s}$. However, the first step can still be expensive when working with thousands of electrons and the knowledge of $N_s$ eigenstates is necessary.

\subsubsection{Sequential exhausting of the full orbital space}
To further compress the $N_{s}$ in the maximization process and, in principle, avoid the knowledge of $N_s$ states altogether, we introduce a \textit{sequential variant} of G-PMWF, sG-PMWF. We first review the approach which assumes $N_s$ states are available, and at the end of this section, we extend it to a more generalized case when the eigenstates do not need to be known \textit{a priori}.

\begin{figure*}
\includegraphics[width=.8\textwidth]{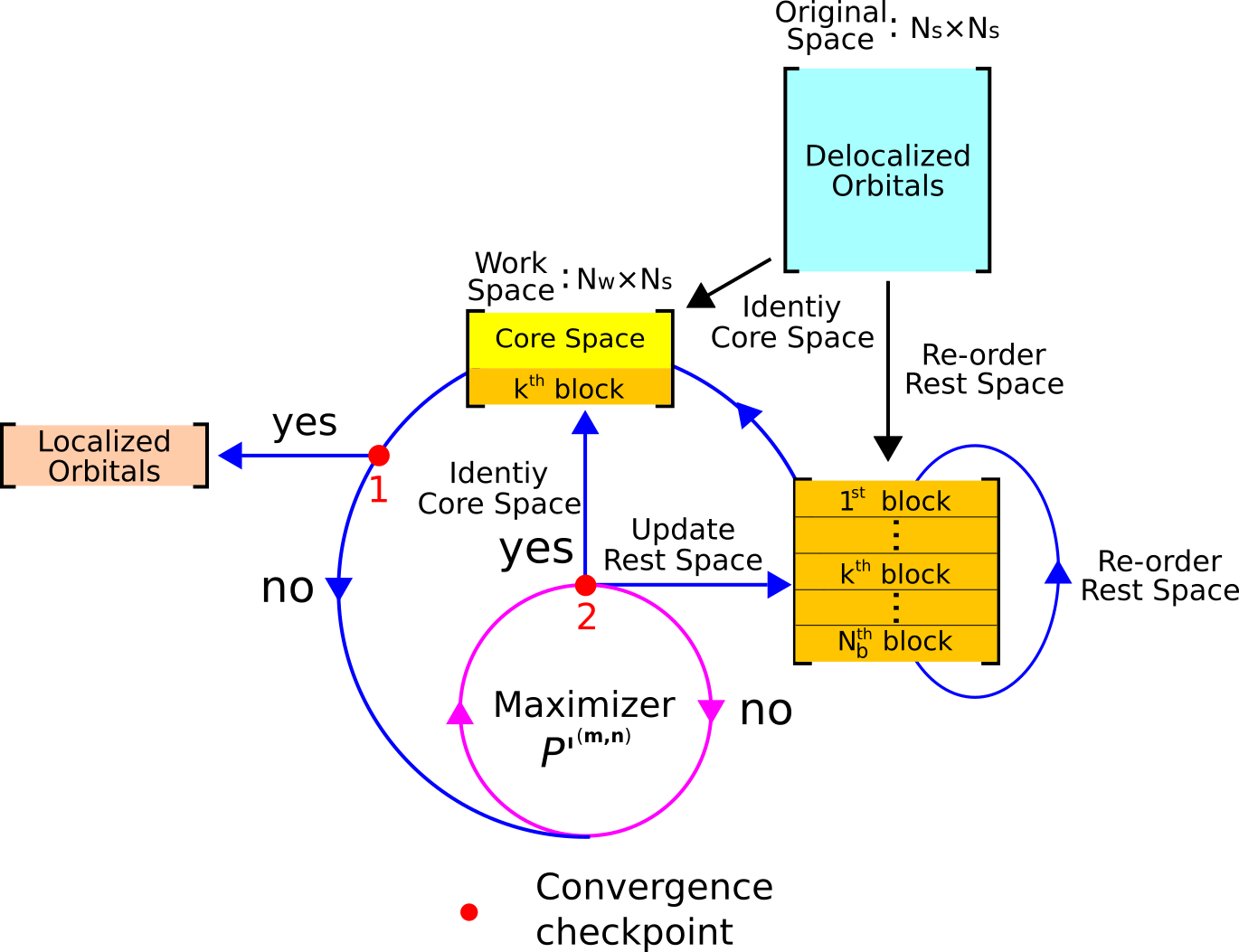}
\caption{Schematic illustration of the sG-PMWF method. Each row of the matrix represents a single-particle state in the canonical $\ket{\phi_j}$ basis. $N_s$ represents the number of states that define the original space while $N_w$ represents the number of states in the actual work space. $P'$ is the modified PM objective functional. The index $m$ denotes the iterative step of the outer-loop (blue), where a different $m$ corresponds to another block that enters the work space. The index $n$ denotes the iterative step of the inner-loop (magenta). If stochastic method is used, the core space will be constructed using random states at $m$=0, and each block of the rest space will be built by mutually orthonormal random states that are complementary to the core space in each $m^{\rm th}$ step.}
\label{fig:fig_sG-PMWF}
\end{figure*}

The sG-PMWF approach incorporates an additional iterative loop (``outer-loop'') to successively maximize the functional $\mathcal{P}'$. The idea is schematically presented in Figure~\ref{fig:fig_sG-PMWF}. A generalized original (entire) space, either occupied or unoccupied, is spanned by $N_s$ orthonormal canonical states. The initial matrix that contains the canonical states is an identity. Each row of the matrix contains the coefficients of a single-particle state in the canonical basis. The number of rows represents the number of states used in the $Q$ matrix. The black lines and arrows stand for the initialization of the localization procedure. The outer-loop is guided by the blue lines and arrows, while the magenta lines and arrows guide the inner-loop (maximizer). The red points denote the convergence checkpoints.

Our goal is to find only $N_{rl}$ states that are spatially localized on a selected fragment, and we seek to minimize the cost of the calculation by neglecting the localization in the other regions of the systems. The general procedure is as follows:

First, we assume that in practical calculations, it may be necessary to account for a ``buffer,'' i.e., we search for $N_c \ge N_{rl}$ states (where $N_c$ is typically similar to $N_{rl}$ in magnitude).  We denote the $N_c$ most localized orbitals chosen based on the value of $L_n^f$ (Eq.~\ref{eq:localityfrag}) as ``core states''. And the ``core space" is spanned by such $N_c$ states. The original space is essentially split into two, the core and its complement space (denoted ``rest space"). The states in the rest space are then reordered upon their locality for the next step.

Second, a work space is built with a dimensionality of $N_w\times N_s$, where $N_c < N_w \ll N_s$. The first part of the work space is filled by the core states (the yellow region). On the other hand, the rest space is partitioned into $N_b$ blocks according to the value of $N_r$, which is an arbitrary number ($N_r\ge$ 1) that denotes the number of states from the rest space. And note that the states in the rest space have been re-ordered in the decreasing order of $L^{f}_i$. The number of states in each block satisfies the following equations
\begin{equation}\label{eq:stblock1}
    N^{k}_{s} = N_{r}  \hspace{5mm} k < N_b,
\end{equation}
and
\begin{equation}\label{eq:stblock2}
    N^{k}_{s} \le N_{r}  \hspace{5mm} k = N_b.
\end{equation}
Here $N_{s}^{k}$ represents the number of states in the $k^{\rm th}$ block. 
The rest space is sequentially updated (explained in the next step) and can be re-accessed during the localization process. The index $m$ denotes and $m^{\rm th}$ iteration step in the outer-loop and the $m$ and $k$ are connected by
\begin{equation}\label{eq:kblock}
k =
\begin{cases}
m \hspace{22.5mm} m\le N_b\\
{\rm mod}(m,N_b) \hspace{5mm} m > N_b.
\end{cases}
\end{equation}

Third, the initial ($m$=0) objective functional value (Eq.~\eqref{eq:Pprime}) is calculated for the work space and the change of the PM functional in the outer-loop is defined as
\begin{equation}\label{eq:deltaPm}
\Delta \mathcal{P}'^{(m)} = \mathcal{P}'^{(m,0)} - \mathcal{P}'^{(m-1,0)} \hspace{5mm} m \ge 1.
\end{equation} 
The convergence checkpoint 1 in Figure~\ref{fig:fig_sG-PMWF} evaluates the $\Delta \mathcal{P}'^{(m)}$ as well as the accumulative step $m$. The iteration will exit the outer-loop if either 
\begin{equation}\label{eq:convergence1}
\Delta \mathcal{P}'^{(m)} \le \text{convergence threshold 1}
\end{equation} 
or
\begin{equation}\label{eq:maxiter1}
m = \text{maximal outer-loop iterations}
\end{equation} 
is satisfied. The convergence threshold 1 and the maximal outer-loop iterations are carefully chosen to converge the localization (see the next section). If the iteration does not exit the loop, the index $m$ will become $m+1$, and the corresponding $k^{\rm th}$ (Eq.~\ref{eq:kblock}) block will fill the second part of the work space. The constructed work space then enters the maximization solver (the inner-loop in magenta). The change of the PM functional in the inner-loop is defined as
\begin{equation}\label{eq:deltaPn}
\Delta \mathcal{P}'^{(n)} = \mathcal{P}'^{(m,n)} - \mathcal{P}'^{(m,n-1)} \hspace{5mm} n \ge 1.
\end{equation}
Here $n$ denotes the iteration step (if iterative maximization is needed) in the inner-loop. The convergence checkpoint 2 evaluates the $\Delta \mathcal{P}'^{(n)}$ as well as the accumulative step $n$. The iteration will exit the inner-loop if either
\begin{equation}\label{eq:convergence2}
\Delta \mathcal{P}'^{(n)} \le \text{convergence threshold 2}
\end{equation} 
or
\begin{equation}\label{eq:maxiter2}
n = \text{maximal inner-loop iterations}
\end{equation} 
is satisfied. The convergence threshold 2 and the maximal inner-loop iterations are carefully chosen to allow the work space to reach the maximum smoothly (see the next section). Once exiting, the core space is identified from the transformed work space, and the residues of work space replace the $N_{s}^{k}$ states in the $k^{\rm th}$ block. This operation is denoted ``the update of the rest space" since both the core and rest spaces are dynamic during the maximization. The index $n$ is reset to 0, and the $\Delta \mathcal{P}'^{(m)}$ arrives at the convergence checkpoint 1. If the iteration does not exit the loop, the next block then fills the work space to re-enter the maximizer. With all the $N_b$ blocks exhausted and updated, the states in the rest space will be re-ordered again for the re-access.

In practice, the outer-loop (identify the core space, construct the work space, maximization, and update the rest space) has to be iterated multiple times until the $\mathcal{P}'$ is converged. In general, each iteration step in the outer-loop feeds the core space with the ingredients to localize itself and sequentially exhaust the full orbital space until convergence. However, the cost of the calculation depends primarily on the size of the work space $N_w$. A small $N_r$ might require extra outer-loop iterations, but the cost of each maximization (``inner-loop'') should be orders of magnitude smaller than the traditional full-space approach. 

So far, we have assumed that a basis of individual single-particle states is known (e.g., obtained by a deterministic DFT). However, this procedure is trivially extended even to other cases, e.g.,  when stochastic DFT is employed\cite{Baer2013,Cytter2018,Chen2019,Fabian2019,Nguyen2021,Baer2022}. For simplicity (and without loss of generality), we assume the localization is performed in the occupied subspace. Here, the sG-PMWF calculation is initialized by constructing a guess of $N_{rl}$ random vectors $\ket{\zeta}$ which are projected onto the occupied subspace as $\ket{\zeta_c} = \hat P^o \ket{\zeta}$. These $N_{rl}$ random states then enter the core space in Figure~\ref{fig:fig_sG-PMWF}. Here, the projector $\hat P^o$ is a low-pass filter constructed from the Fermi operator leveraging the knowledge of the chemical potential\cite{Baer2013,Cytter2018,Chen2019,Fabian2019,Nguyen2021,Baer2022,neuhauser2014breaking,Vlcek2018swift}. Next, in each outer-loop step, one creates a block of random vectors $\ket{\zeta_r^m}$ which have to be mutually orthogonal as well as orthogonal to the $N_{rl}$ core states via, e.g., Gram-Schmidt process. Here $r$ denotes the rest space and $m$ denotes the $m^{\rm th}$ step in the outer-loop. This block of random states follows the procedure in Figure~\ref{fig:fig_sG-PMWF} to fill the work space. Note that this block of random vectors represents the entire orthogonal complement to the core space.

Combined with the fragmentation treatment, the number of elements in $Q$ is reduced from $N_A \times N_s^2$ to $N_w^2$. And the unitary matrices are also reduced from $N_s^2$ to $N_w^2$. Such a reduction in dimensionality is expected to shorten the time spent on each iteration step as long as $N_w \ll N_s$. The cost of the stochastic method (which does not require the knowledge of the $N_s$ eigenstates) is slightly higher due to the additional orthogonalization process. In the Results and Discussion section, we show that the total wall time spent on a job becomes much shorter, especially for giant systems, at the expense of more inner-loop steps. Most importantly, the localized states obtained from sG-PMWF are practically identical to those obtained from the traditional G-PMWF approach.

\section{Computational Details}

\subsection{G-PMWF and sG-PMWF}
A shared memory approach is employed to parallelize the do-loops (via  OpenMP). Hirshfeld partitioning\cite{Hirshfeld1977} is adopted for the atomic weight function in calculating the $Q$ matrix. For simplicity, we employ the steepest ascent (SA) algorithm\cite{Silvestrelli1998,Silvestrelli1999,Thygesen2005} to maximize the PM functional $\mathcal{P}$ and $\mathcal{P}'$. Note that other extremization procedures will likely further reduce the cost of the inner-loop, but they do not have a decisive effect on the overall scaling.  The ascending step is set 5.0 at the beginning and is divided by 1.1 each time the change of PM functional $\Delta \mathcal{P}'^{(n)}$ appears negative. The stochastic basis in the sG-PMWF calculations is constructed using Fortran random number generator. The random number generator employs seeds that change in each outer-loop step. 

In traditional G-PMWF calculations, the convergence criterion is set $<5 \times 10^{-7}$ and, it has to be consecutively hit three times to ensure smooth convergence. In sG-PMWF calculations, the convergence criterion is set $10^{-7}$ (convergence threshold 2) in the inner-loop, which also has to be hit three times consecutively. The convergence threshold 1 is set $5 \times 10^{-7}$ for the outer-loop. The maximal iteration steps is set 2000 for $n$ and 5000 for $m$.

To avoid the spurious convergence or local maximum issue, a special criterion is devised for the sG-PMWF. The principle comes from the traditional G-PMWF. When the core space reaches the maximum localization, the whole rest space should no longer increase the $\mathcal{P}'$ by $>5\times 10^{-7}$ and neither should a subspace in the rest space contribute further. And thus, the $\Delta \mathcal{P}'^{(m)}$ of each block in one complete access of the rest space are evaluated simultaneously. Only the maximal $\Delta \mathcal{P}'^{(m)}$ satisfies the criterion ($<5\times 10^{-7}$) will the $\mathcal{P}'^{(m)}$ be considered converged. This also means, once the $1^{\rm st}$ block re-enters the work space, all the blocks have to be exhausted to decide the convergence. This might lead to a slight increase in cost but guarantees that the sG-PMWF reaches the convergence in the same manner as the G-PMWF does.

The sG-PMWF calculation can be easily restarted as long as one keeps the checkpoint file at the previous $m^{\rm th}$ step and sets the outer-loop to start with $m+1$. The source code is posted on git-hub and available for download.

\subsection{Model systems}
As a test case, we investigate the negatively-charged nitrogen-vacancy (NV$^-$) center in 3D periodic diamond supercells and a 2D slab. The atomic relaxations of the NV$^-$ defect center in 3D periodic diamond supercells with 215, 511, and 999 atoms are performed using QuantumESPRESSO package\cite{QE2017} employing the Tkatchenko-Scheffler's total energy corrections\cite{TS_2009}. For the 111 nitrogen terminated surface slab 2D periodic calculations, the surface relaxation also employs the Effective Screening Medium correction\cite{Otani_2006}. The atom relaxation of the surface terminated with nitrogen atoms is performed on a smaller slab with 24 atoms, which corresponds to the $1\times1\times2$ supercell. The relaxed top and bottom surfaces were then substituted into a large $4\times4\times6$ ($1.5\times1.7\times4.7$~nm) supercell containing 2303 atoms. The 111 surface is set normal to the $z$-direction. The relaxed structure of the NV$^-$ center is cut out from a 511 3D periodic supercell in a way that the N--V axis is normal to the 111 surface. This supercell is then substituted in the middle of the 111 nitrogen terminated surface $4\times4\times6$ slab at the 2~nm depth from the surface. 

The starting-point calculations for all systems are performed with a real-space DFT implementation, employing regular grids, Troullier-Martins pseudopotentials\cite{TroullierMartins1991}, and the PBE\cite{PerdewWang} exchange-correlation functional. For 3D periodic structures, we use a kinetic energy cutoff of 26 Hartree to converge the eigenvalue variation to $< 5$~meV. The real-space grids of $68\times 68 \times 68$; $92\times 92 \times 92$; and $ 112 \times 112 \times 112$ with the spacing of 0.3~$a_0$ are used for 215-atom, 511-atom, and 999-atom supercells, respectively; The grid of $70\times 82 \times 338$ with the spacing of 0.4~$a_0$ is used for 2303 atoms slab supercell. The generated canonical Kohn-Sham eigenstates are used for the subsequent orbital localization.

\section{\label{sec:result}Results and Discussion}
The traditional G-PMWF and the proposed sG-PMWF methods are applied to obtain regionally localized states on the NV$^-$ center in diamond. The NV$^-$ center is composed of three carbon atoms and one nitrogen atom that are mutually non-bonded. The fragment in the actual calculations is constructed with these four atoms (see Figure~\ref{fig:fig_fragment}a) unless stated otherwise. The number of regionally localized states, $N_{rl}$, is 16 on the constructed fragment. Two types of systems, solids and slab, are studied. For the solids, three supercells of different sizes are investigated. The number of occupied states, $N_s$, for each system is 432, 1024, and 2000, respectively. For the slab, the regionally localized states are identified from a supercell with 2303 atoms and 4656 occupied states.

\subsection{Completeness of sG-PMWF}
First, we investigate the completeness of the sequential exhausting approach, i.e., whether the sG-PMWF can reproduce the same results as the G-PMWF. To contrast the sG-PMWF method, we perform G-PMWF localization on the 511-atom system using a truncated orbital space. This is a common technique to lower the cost by filtering out a portion of canonical states upon the eigenenergy (eigenvalue). Only eigenstates within a specific energy range (termed as the ``energy window") are selected for localization. We tested two energy windows (10 eV and 20 eV below the Fermi level, respectively) on obtaining the localized Wannier function basis. Upon visual inspection, the results do not look too different, but when applied to compute the optical transitions in the NV$^-$ center (see ``Downfolded effective Hamiltonian" in the SI), we see considerable differences in the energies (Table~\ref{tab:table_window}). The results from the truncated space are highly underestimated compared with the results from the full space. The energy-windowing technique fails since, to reach optimal localization, the maximum possible Bloch states are needed to be transformed, i.e., \textit{all} the occupied states are necessary. The proposed sG-PMWF method does not suffer from this problem, and we demonstrate its completeness below.

\begin{figure}
    \centering
    \includegraphics[width=3.37in]{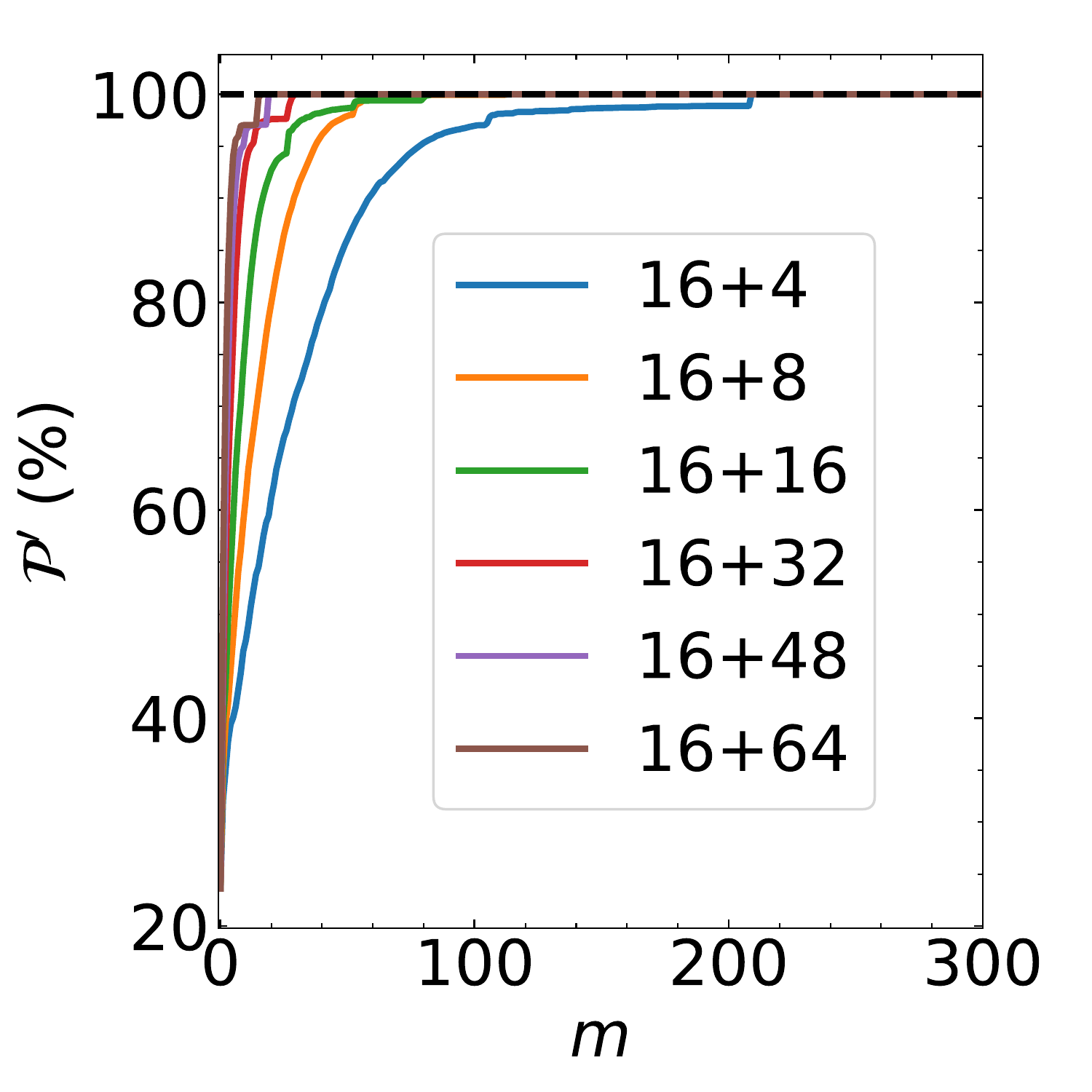}
    \caption{Convergence of the functional $\mathcal{P}'$ with respect to the outer-loop step $m$ for the NV$^-$ center of the 215-atom system. Each curve is labelled by the combination of $N_c$ and $N_r$.}
    \label{fig:fig_steps_P_obj}
\end{figure}

\begin{figure}
    \centering
    \includegraphics[width=3.37in]{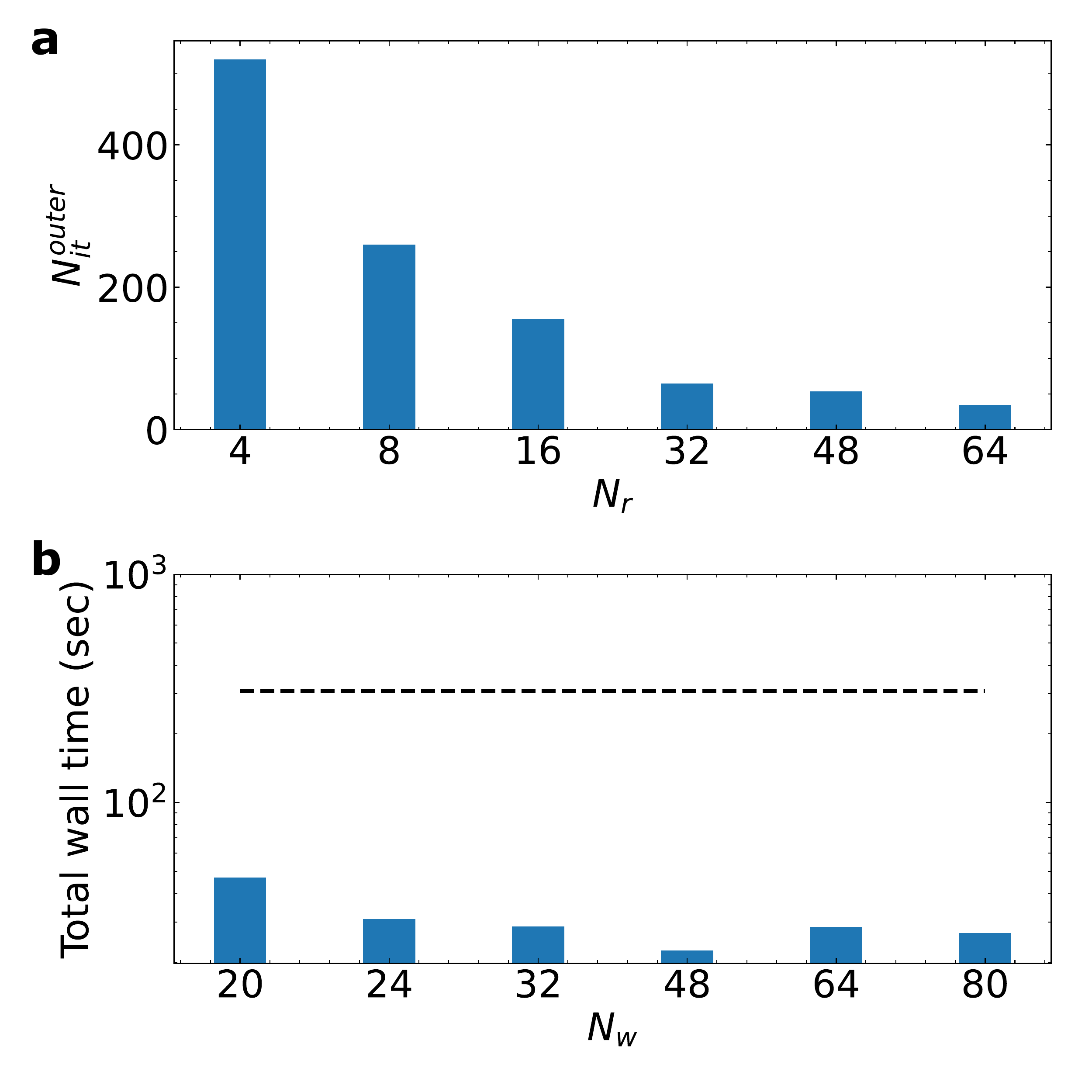}
    \caption{Investigation of different combinations of $N_c$ and $N_r$ for the localization on the NV$^-$ center of the 215-atom cell. $N_c$ is fixed at 16. (a) Number of iteration steps in the outer-loop as a function of the $N_r$. (b) Total wall time of the calculation as a function of $N_w$. The dashed line indicates the total wall time from the G-PMWF method using the full orbital space.}
    \label{fig:fig_steps_Nr}
\end{figure}

We first illustrate the completeness in detail using the 215-atom system.  To initialize the sG-PMWF calculations, the $N_c$ parameter takes 16 (minimum), i.e., we take no ``buffer." For convenience, we only consider combinations with $N_r$ being an integer multiple of $N_c$ and vice versa. Several $N_r$ ranging from 4 to 64 are tested. Figure~\ref{fig:fig_steps_P_obj} shows the maximized $\mathcal{P}'$, which measures the degree of localization (Eq.~\eqref{eq:Pprime}), relative to the converged maximized value using the full space ($\mathcal{P}' / \mathcal{P}'_{full}$) as a function of the accumulative outer-loop step $m$. It can be clearly seen that 100$\%$ of the $\mathcal{P}'_{full}$ is sequentially recovered regardless of the ($N_c,N_r$) combination. The maximization of each curve presented in Figure~\ref{fig:fig_steps_P_obj} is not smooth, i.e., spikes are observed at the step where the dynamic rest space is re-accessed. Table~\ref{tab:table_215_tt} in the supporting information (SI) shows that at least 94$\%$ of the converged $\mathcal{P}'$ has been gained after completing the first access. As the $N_r$ increases, fewer and fewer iteration steps ($N_{it}^{outer}$) are required to reach convergence (Figure~\ref{fig:fig_steps_Nr}a). And theoretically, the $N_{it}^{outer}$ should be reduced to two (the second step is to exit the outer-loop) if one takes $N_r = N_s - N_c$ to work directly in the full space. The reduction in $N^{outer}_{it}$, however, does not necessarily lead to a shorter job time. Note that the time per outer-loop iteration ($t^{outer}$) increases with a scaling of $\mathcal{O}(N_w^{1.53})$ (see Figure~\ref{fig:fig_scaling_tpi}) for the 215-atom system. Figure~\ref{fig:fig_steps_Nr}b shows the total wall time of each job as a function of the $N_w$ with $N_c$ fixed at 16. The $N^{outer}_{it}$ dominates the total wall time when $N_w$ is small ($<$48). In this regime, reducing the number of iterations lowers the total wall time effectively. When the $N_w$ is larger, however, the $t^{outer}$ becomes the dominating factor and the total wall time increases even though the $N^{outer}_{it}$ decreases. The trade-off between $N^{outer}_{it}$ and $t^{outer}$ suggests there exists an optimal combination of $N_c$ and $N_r$ for a specific system to minimize the total cost.

In Table~\ref{tab:table_215_tt}, the last row shows the sG-PMWF calculation employing a set stochastic basis that represents the rest space. The same parameter combination (16,32) is used. The 16 core states are taken directly from the canonical eigenstates based on the locality, while the 32 stochastic states are constructed in a three-step manner (see ``Preparation of stochastic basis" in the SI). Compared with the (16,32) calculation using the deterministic basis, the stochastic approach exhibits the same completeness in exhausting the full orbital space, as seen from the converged $\mathcal{P}'$ and $\mathcal{P}$. Nevertheless, more outer-loop iterations are needed due to the randomized search. And the time per iteration also becomes longer (3.47 seconds versus 0.32 seconds) due to the Gram-Schmidt orthogonalization process. And therefore, the total wall time increases to 729 seconds. Figure~\ref{fig:fig_steps_P_obj2} shows the evolution of the objective function as a function of the outer-loop step $m$. In comparison with the deterministic counterpart, the stochastic approach converges more smoothly. The stochastic basis search does not show competitive efficiency versus the full-space approach (308 seconds) for such a small system. In the following section, we show the stochastic basis approach becomes more efficient than the full-space counterpart for a larger system. However, we emphasize that the advantage of sG-PMWF does not hinge on this stochastic extension, but it enables it. In most of our results, we will focus on the fully deterministic approach in which the knowledge of $N_s$ states is assumed.

The behavior of the sG-PMWF method discussed above is also observed for the 511-atom system (Figure~\ref{fig:fig_steps_Nr_511} and Table~\ref{tab:table_511_tt} in the SI), confirming the generality of the completeness.

\subsection{Optimization of Work Space}
\begin{figure}
    \centering
    \includegraphics[width=3.37in]{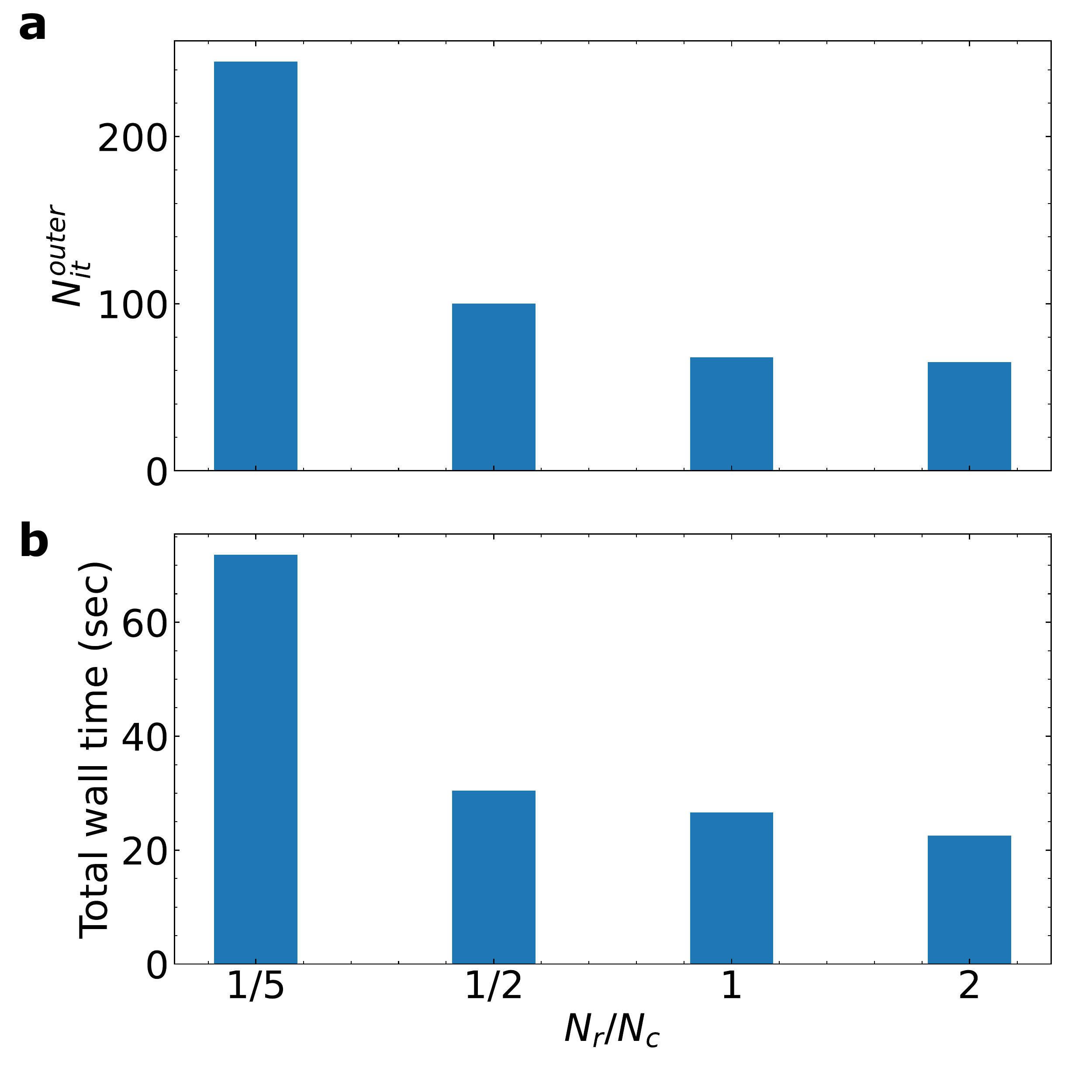}
    \caption{Investigation of different combinations of $N_c$ and $N_r$ for the localization on the NV$^-$ center of the 215-atom cell. $N_w$ is fixed at 48. (a) Number of iteration steps in the outer-loop as a function of the $N_r/N_c$ ratio; (b) The total wall time as a function of the $N_r/N_c$ ratio.}
    \label{fig:fig_Nr_Nc_wall}
\end{figure}

In the previous section, we observe a trade-off between $N^{outer}_{it}$ and $t^{outer}$, which implies a possibly optimal parameter combination. To further understand the choices of $N_c$ and $N_r$, several other combinations with $N_c > 16$ are tested on the 215-atom system, with results summarized in Table~\ref{tab:table_215_tt}. The maximal $\mathcal{P}'$ and $\mathcal{P}$ are secured regardless of the ($N_c,N_r$) combination, indicating that the convergence of $\mathcal{P}'$ is insensitive to the choices of these two parameters. For $N_c$ fixed at 16, the time-to-solution reaches a minimum when $N_w=48$, as shown in Figure~\ref{fig:fig_steps_Nr}. For $N_w$ fixed at 48, different ratios of $N_r/N_c$ are tested, which turns out the larger the $N_r$, the smaller the $N^{outer}_{it}$ (Figure~\ref{fig:fig_Nr_Nc_wall}a). Note that the $t^{outer}$ depends solely on the $N_w$ (Table~\ref{tab:table_215_tt}). And therefore a smaller $N^{outer}_{it}$ translates directly to a shorter wall time (Figure~\ref{fig:fig_Nr_Nc_wall}b). This behavior is further observed in the 511-atom system (see Figures~\ref{fig:fig_Nr_Nc_wall_511} in the SI). 

To conclude, the ``buffer" seems to be unnecessary for the core space, i.e., $N_c$ can be set directly as $N_{rl}$ for a specific fragment. The work space optimization then depends solely on the choice of $N_r$. Nevertheless, the cost of the investigated sG-PMWF calculations without optimization is already absolutely lower than that of G-PMWF regardless of the $N_r$ (see Figure~\ref{fig:fig_steps_Nr}b and Figure~\ref{fig:fig_steps_Nr_511}b). The protocol of choosing $N_c$ and $N_r$ is suggested to be $N_c = N_{rl}$ and $N_r =2N_c$ since it leads to a local minimum in the total wall time. 

This protocol is then applied to the 999-atom system and two additional combinations of $N_c$ and $N_r$ are also tested (Table~\ref{tab:table_999_tt}). The (16,32) combination still leads to a cost minimum and is 85 times faster than the G-PMWF. Further, we also test the stochastic basis search with the 999-atom employing the (16,32) combination (see the last row of Table~\ref{tab:table_999_tt}). The completeness of the stochastic exhausting is again confirmed by the converged $\mathcal{P}'$ and $\mathcal{P}$. Although the stochastic approach is still more costly than the deterministic sequential counterpart, it is more efficient than the full-space G-PMWF calculation (by roughly 50\%) when applied to this system with $\sim$4000 electrons. Furthermore, $\sim$74\% of the cost in the stochastic search comes from the Gram-Schmidt process, which advanced orthogonalization techniques can optimize. When combined with stochastic DFT, the total cost of orbital localization is expected to be much lower than the deterministic approach that requires the knowledge of the eigenstates in a system with tens of thousands of electrons.

For the 2303-atom system (Table~\ref{tab:table_2303_tt}), the (16,32) combination successfully converges the $\mathcal{P}'$ and produces localized states. Note that the cost can be lowered by 10\% if the (16,48) combination is used. And if one searches further for the optimal $N_r$ (or $N_w$), it is possible to lower the cost further. However, for a fair comparison between one system and another, we use the timing from the (16,32) combination for the slab, which is already 412 times faster than the G-PMWF.

In Table~\ref{tab:table_Pptime}, we compare the time spent on maximizing the $\mathcal{P}'$ (Eq.~\eqref{eq:Pprime}, the folding step) and maximizing the $\mathcal{P}$ (Eq.~\eqref{eq:p_2}, the unfolding step). In each system, the cost of the unfolding step is merely 1$\sim$2\% of the folding one since only $N_{rl}$ states are transformed in the unfolding step. And thus, it is sufficient to evaluate just the cost of the unfolding step as the total cost of the orbital localization.

Finally, we remark that the (16,32) combination is stable and efficient for a given fragment regardless of the precise environment. This indicates that sG-PMWF is robust. Further, the consistent parameter combination clearly demonstrates the scaling of the sG-PMWF calculation with respect to the $N_s$ as discussed in the next section.

\subsection{Scaling analysis of sG-PMWF vs. G-PMWF}
To investigate the scaling of the sG-PMWF method, we first study the scaling of the time per outer-loop step ($t^{outer}$) with respect to the $N_s$. The $t^{outer}$ is normalized to the largest grid
\begin{equation}
    t^{outer}_n = \frac{N_{g}^{max}}{N_{g}} t^{outer},
\label{eq:tnorm}
\end{equation}
where $t^{outer}_n$ represents the normalized time step, $N_{g}^{max}$ denotes the number of grid points of the largest system (2303-atom system), and $N_{g}$ is the grid of each investigated system. It can be shown that the $t^{outer}_n$ is independent of the number of occupied states ($N_s$) in the system (Table~\ref{tab:table_tpi_1}) when the same ($N_c,N_r$) combination is applied. This $N_s$-independence implies a possibly lower scaling in the total computational cost compared with G-PMWF.

\begin{figure}
    \centering
    \includegraphics[width=3.37in]{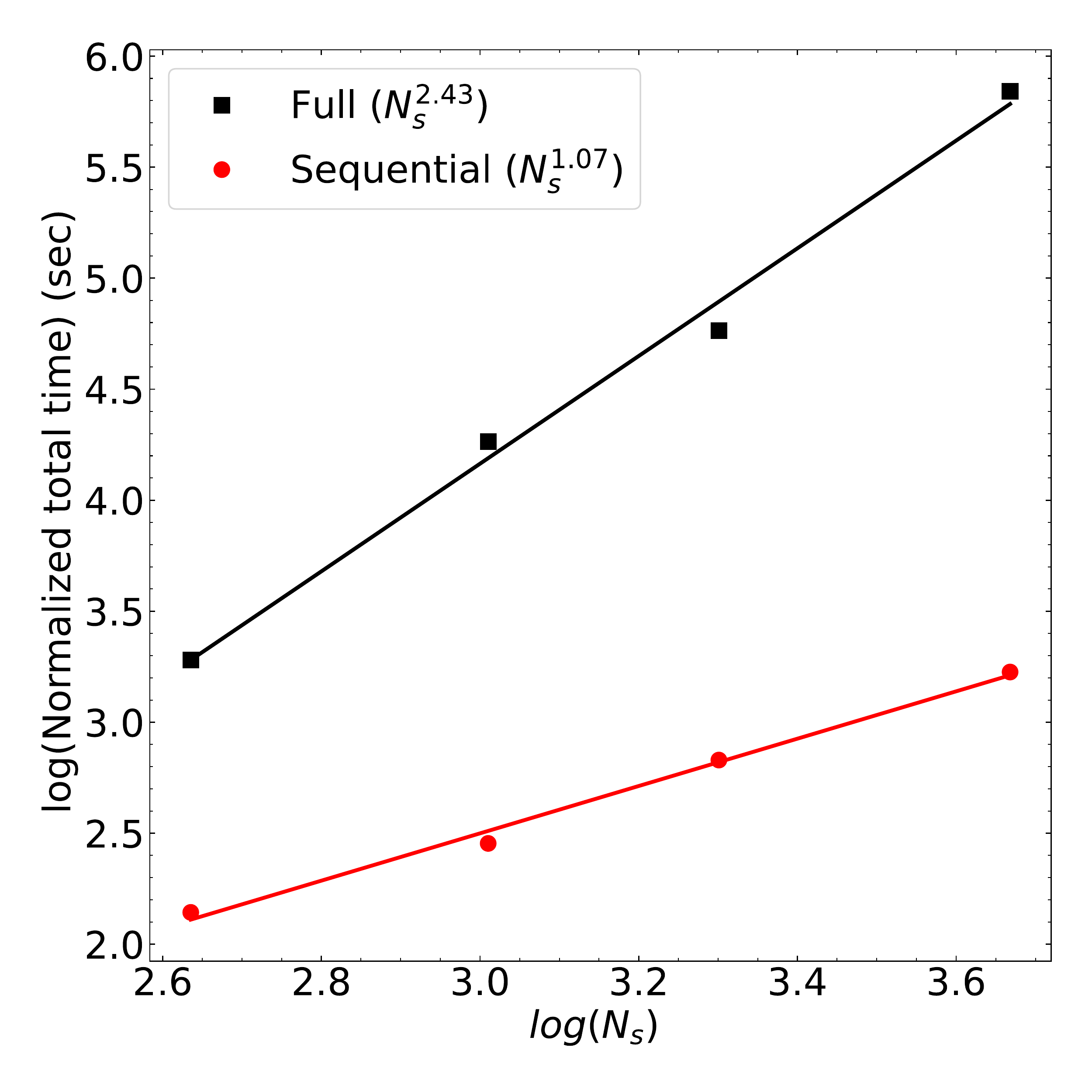}
    \caption{The log of the normalized total job time plotted as a function of the log of $N_s$ for the four investigated systems. The black line and square points represent results obtained from the G-PMWF method using the full orbital space. The red line and circle points represent results obtained from the sG-PMWF method using the constructed work space. The total job time is normalized to the largest grid (2303-atom system). The scaling is derived from the slope of each fitting using the numeric data in Table~\ref{tab:table_total_norm}.} 
    \label{fig:fig_wall_time_Ns}
\end{figure}

In Figure~\ref{fig:fig_wall_time_Ns}, the log of the normalized total job time (similar to Eq.~\eqref{eq:tnorm}) is plotted as a function of the log of $N_s$ for the four investigated systems. The scaling of the G-PMWF using the full orbital space is $\mathcal{O}(N_s^{2.43})$ (black line and square points). This is a bit higher than the theoretical $\mathcal{O}(N_s^{2})$ due to the other $\mathcal{O}(N_s)$ do-loops, tasks related to parallelization, and practical executions (e.g., reading and writing of files). The sequential method, sG-PMWF, reduces the scaling from $\mathcal{O}(N_s^{2.43})$ to $\mathcal{O}(N_s^{1.07})$ (read line and circle points). This linear scaling is observed when the same protocol (16,32) applies to the four systems. Such an order of magnitude reduction in the scaling promises the efficiency of sG-PMWF when applied to much larger systems. In our largest system with 4656 states, the total wall time is shortened from 8 days to $<$ 0.5 hour (on a work station with 2.5 GHz CPUs and parallelization on 60 cores)  (Figure~\ref{fig:fig_total_wall_time}).

The reduced scaling of sG-PMWF is largely attributed to the reduction of dimensionality during the maximization process. The efficiency is reflected mainly in the time per inner-loop iteration, $t^{inner}$ (Table~\ref{tab:table_tpi_2}). From 432 states to 4656 states, the $t^{inner}$ of the G-PMWF approach scales rapidly from 0.29 seconds to 1056 seconds. In sG-PMWF, however, the $t^{inner}$ remains constant and extremely low ($\sim5\times 10^{-4}$ seconds) regardless of the $N_s$. Although more SA iterations steps are required relative to the G-PMWF calculations (Figure~\ref{fig:fig_iter_Nr_215} and~\ref{fig:fig_iter_Nr_511}), 1000 iterations now take as low as 0.5 seconds. And therefore, in sG-PMWF, the time spent in the maximizer is no more the dominating factor within an outer-loop step. It is sufficient to evaluate the efficiency of sG-PMWF by the $t^{outer}$. As shown in Figure~\ref{fig:fig_time_per_iter}, the scaling of the normalized $t^{inner}$ (similar to Eq~\eqref{eq:tnorm}) in G-PMWF is $\mathcal{O}(N_s^{2.64})$, while the $t^{outer}_n$ in sG-PMWF hardly scales with $N_s$. Further, Table~\ref{tab:table_Nit} shows that the numbers of inner-loop iterations $N_{it}^{inner}$ in G-PMWF are reasonably large (600$\sim$700) and translate to a total scaling of $\mathcal{O}(N_s^{2.43})$ shown in Figure~\ref{fig:fig_wall_time_Ns}. In sG-PMWF, the $N_{it}^{outer}$ scales almost linearly with $N_s$ and gives a total scaling of $\mathcal{O}(N_s^{1.07})$. 

\subsection{Localization quality of sG-PMWF vs. G-PMWF}
\subsubsection{Visualization of localized orbitals and density}
\begin{figure}
    \centering
    \includegraphics[width=3.37in]{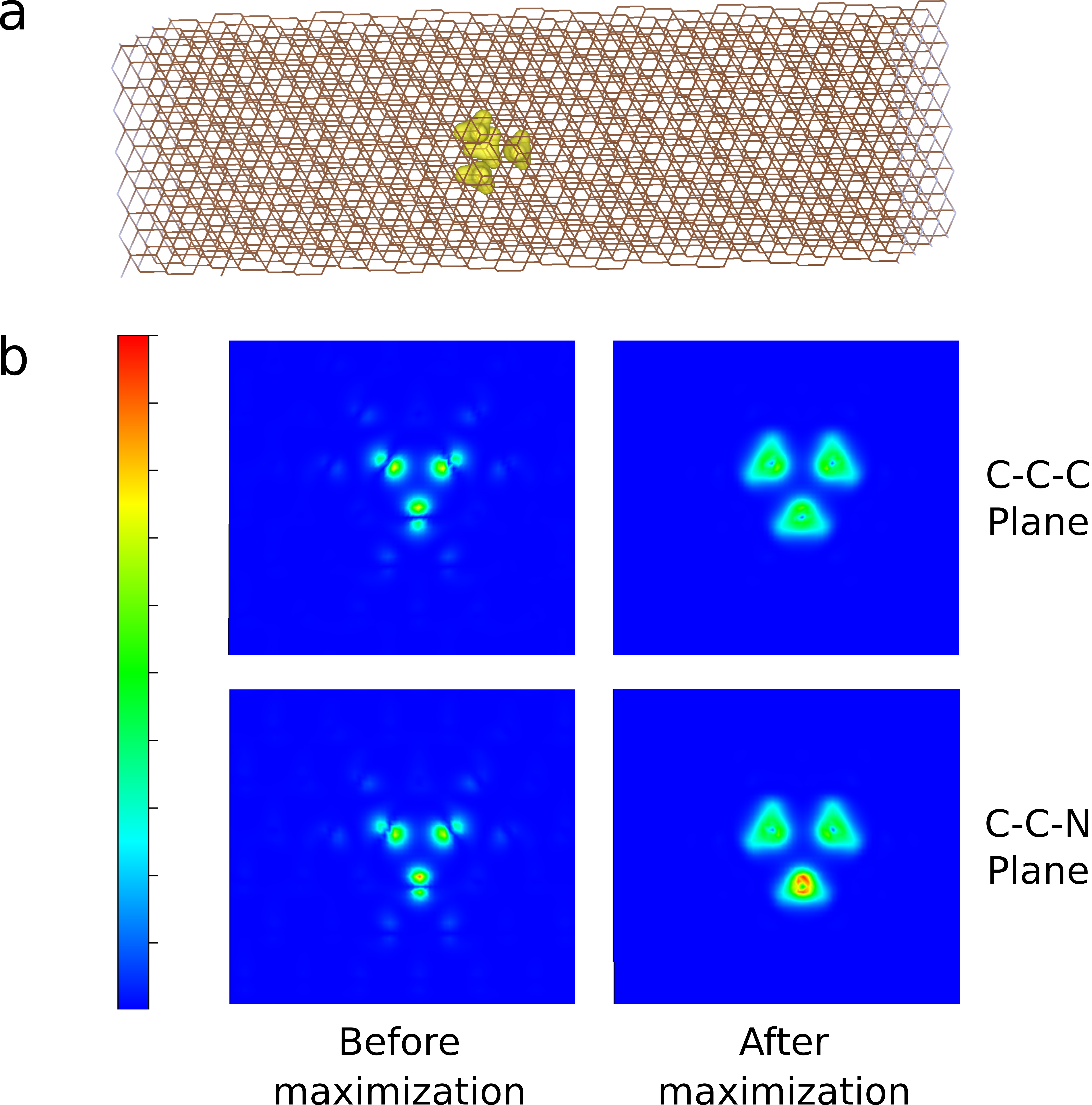}
    \caption{(a) Electron density constructed from the 16 regionally localized states around the NV$^-$ center. The isosurface value is set 0.05. (b) density distribution sliced through the C-C-C plane (upper panels) and the C-C-N plane (lower panels) of the NV$^-$ center in the slab. The left panels are constructed from the 16 most localized canonical states before the sG-PMWF calulation and the right panels are constructed from the 16 regionally localized states after the sG-PMWF maximization.}
    \label{fig:density2D_plot}
\end{figure}

In the previous section, the completeness of sG-PMWF has been demonstrated for the maximization of the modified PM functional $\mathcal{P}'$ (Eq.~\ref{eq:Pprime}). These 16 resulting states are localized on the fragment and serve as a subspace to further maximize the $\mathcal{P}$ (Eq.~\ref{eq:p}), which unfolds the states on each individual atom. Table~\ref{tab:table_cmpp} and~\ref{tab:table_cmp} summarize the converged $\mathcal{P}'$ and $\mathcal{P}$. The results of the sG-PMWF method differ from that of G-PMWF by no more than 0.0001 ($<0.002\%$). Figure~\ref{fig:fig_dens215} to~\ref{fig:fig_dens999} show that the electron density constructed from the 16 regionally localized states are visually identical between the sG-PMWF and G-PMWF calculations. The same agreement is also seen for the four selected individual ``p-like" states (Figure~\ref{fig:fig_rls215} to~\ref{fig:fig_rls999}). Figure~\ref{fig:density2D_plot}a highlights the NV$^-$ center in the slab using the regionally localized electron density. The obtained electron density conserves the spatial symmetry across the C-C-C plane and the C-C-N plane (Figure~\ref{fig:density2D_plot}b). The left panels of Figure~\ref{fig:density2D_plot}b show the electron density constructed from the 16 most localized canonical states, while the right panels present the maximized results from the sG-PMWF calculation. It can be clearly seen that electron density distribution becomes much more concentrated on the selected atoms, indicating the effectiveness of the localization.

To demonstrate that the sG-PMWF localization is subsystem-independent, an arbitrary carbon atom is chosen from each investigated system, and four regionally localized states are sought. Figure~\ref{fig:fig_1cdens} shows that the electron density around the selected C atom is successfully reproduced for each system, confirming the generality of the sG-PMWF approach.

\subsubsection{Excited states of the NV$^-$ center}
To further demonstrate the practical application and quality of the sG-PMWF approach, we investigate the optical transitions in the NV$^-$ center using the ``p"-like Wannier function basis (see Figure~\ref{fig:fig_rls215} to~\ref{fig:fig_rls999}). To model the excited states of the NV$^-$ center, we solve the Hubbard Hamiltonian, defined in Eq.~\eqref{eq:hamiltonian} in the SI. It is a minimal model of the NV$^-$ center that is commonly used~\cite{choi2012mechanism,bockstedte2018ab,ma2020quantum,ma2021quantum} to describe its low-lying excited states. In this section, we will, in particular, comment on the selection of the fragment on which the electronic states are localized. Note that the fragment size is independent of the sG-PMWF methodology, but it represents an important parameter. 

First, we focus on the results computed from the localized basis on the four-atom fragment (Figure~\ref{fig:fig_fragment}a). The results for the three lowest energy transitions are in Table~\ref{tab:table1} in parentheses. For the 3D periodic systems, the $^3E$ -- $^3A_2$ transition energy is slightly underestimated, while the $^1A_1$ -- $^3A_2$ one is overestimated in the two small cells. For the bulk systems, the $^3E$ -- $^3A_2$ and $^1A_1$ -- $^1E$ transition energies are underestimated with respect to the experimental values of $1.95$~eV and $1.19$~eV, respectively. These results, however, are in good agreement with other theoretical calculations that employ PBE functionals to compute the bare Hubbard model parameters\cite{Goss1996,Gali2008,Delaney2010,Ma2010,Gordon2013,Alkauskas2014a}. The $^1E$ -- $^3A_2$ transition energy fluctuates mildly with respect to the supercell size but maintains a comparable magnitude. The results computed from the sG-PMWF basis agree perfectly with the G-PMWF ones (see Table~\ref{tab:ese_full} in the SI), confirming the equivalency of the two sets of localized orbitals. 

Despite this, the slab results are strikingly different. Again, G-PMWF and sG-PMWF perfectly agree, but the transition energies are up to 70\%-80\% lower than those in the bulk. As we show below, this is due to the selection of the fragment size and independent of the completeness of the orbital space. To the best of our knowledge, we note that no calculations for shallow NV$^-$ centers in slabs have been done previously. Hence it is not possible to compare our results with any reference.

\begin{table}[H]
\centering
\caption{Excited-state transition energies of the NV$^-$ center in the four investigated systems using the Wannier function basis obtained from sG-PMWF calculations. The numbers with and without the parenthesis correspond to the \{4,4\} and \{16,16\} fragment, respectively.}
\label{tab:table1}
\resizebox{\textwidth}{!}{\begin{tabular}{ccccc}
\toprule
\multirow{2}{*}{\begin{tabular}[c]{@{}c@{}}Transition\\ symmetry\end{tabular}} & \multicolumn{4}{c}{Energy (eV)}                               \\ \cline{2-5}
                                                                               & 215-atom cell & 511-atom cell & 999-atom cell & slab          \\ \midrule
$^3E$ -- $^3A_2$                                                                         & 2.108 (1.560) & 2.277 (1.695) & 2.312 (1.710) & 1.343 (0.363) \\
$^1A_1$ -- $^3A_2$                                                                        & 1.433 (1.325) & 1.310 (1.270) & 1.202 (1.193) & 1.159 (0.292) \\
$^1E$ -- $^3A_2$                                                                         & 0.447 (0.378) & 0.435 (0.381) & 0.413 (0.368) & 0.329 (0.091) \\ \bottomrule
\end{tabular}
}
\end{table}

The slab results become more in line with bulk values if a larger fragment size is employed. The fragment studied in the previous sections is in fact a minimal model, i.e., the orbital localization is considered only on the 4 atoms where the ``p"-likes states are located and, the total number of orbitals on these four atoms is 16. However, neglecting the neighboring atoms might lead to a mixed character of ``p"-like states and C-C (or N-C) covalent bonds. To test this, we investigate four combinations of \{$N_A'$,$N_A''$\} fragments: for instance, \{4,16\} represents the case where 4 atoms are considered in maximizing $\mathcal{P}'$ (Eq.~\eqref{eq:Pprime}) while 16 atoms (including the bonded atoms) are considered in maximizing $\mathcal{P}$ (Eq.~\eqref{eq:p_2}). A detailed investigation of the various parameters is performed on the 215-atom system. The corresponding fragments are presented in Figure~\ref{fig:fig_fragment}. The four Wannier functions used for the Hubbard model are illustrated in Figure~\ref{fig:fig_pstates}. 

Numerically, the \{4,16\} combination gives the closest solutions to the result computed using G-PMWF with localization on all atoms at once. We call these results ``all-atom'' calculations. Note that in this case, the optimization does not preferentially localize single-electron states near the defect; rather it seeks globally most localized states. Such an approach is not guaranteed to generate transformed PMW orbitals that are optimal for the mapping onto the Hubbard model. Indeed, we show this numerically below. For a better comparison, we also provide the spatial overlaps between the fragmentation approaches and the all-atom calculation, $\lvert \braket{\psi_i | \psi_j} \rvert$ in Table~\ref{tab:spovlp_atoms}.

In contrast, the results for the \{4,4\} combination represent the minimal fragment where the optimization is performed for sixteen orbitals on four atoms neighboring the defect center. These minimal PMWFs are showin in Figure~\ref{fig:fig_pstates} and displays over-localization of the ``p"-like states in the NV$^-$ center, i.e., the orbitals are less centered on the atoms and tend to merge at the geometric center. This is a purely numerical artifact of a too small optimization space which is alleviated (Figure~\ref{fig:fig_pstates}) when the 12 bonded atoms are included to compete with the geometric center for the electron density. Due to this, we disregard the \{4,4\} case further. 

Upon visual inspection, the \{16,16\} combination \textit{graphically} gives the most localized ``p"-like orbitals (the third row in Figure~\ref{fig:fig_pstates}). To provide a quantitative measure of localization, we calculate the locality of each ``p"-like state on the corresponding atom plus its neighboring bonded atoms to account for the environment
\begin{equation}\label{eq:wflocality}
  L_i = \sum_{A=1}^{4}\braket{\psi_i | w_A |\psi_i},
\end{equation}
where $i$ denote the $i^{\rm th}$ ``p"-like state and $A$ sums over the four 4 atoms (1 atom + 3 bonded atoms). The value for each individual state is summarized in Table~\ref{tab:loc_atoms} where we use the sum, $\sum_{i=1}^4 L_i$, to represent the whole set of PMW functions. In agreement with the visual analysis, the \{16,16\} combination exhibits the strongest localization (Figure~\ref{fig:fig_pstates}) attributed to the modification of the objective functional (Eq.~\ref{eq:Pprime}). As commented by J$\acute{\text{o}}$nsson\cite{Jonsson2017} et al., the solutions to ``maximally-localized Wannier functions" are actually not unique and sometimes ambiguous since the resulting localized orbitals are determined by the objective functionals. We emphasize that the traditional G-PMWF approach evaluates the overall orbital localization on all the atoms, but it does not necessarily reach maximal localization on a specific subsystem (fragment). Instead, the proposed fragmentation treatment in this work leads to an objective functional for regionally localized orbitals. We surmise that this approach is more beneficial for effective embedding and downfolding.

To further analyze the results, we use the four sets of PMW functions and compute the optical transition energies for the 215-atom system (Table~\ref{tab:table2}). The \{4,16\} combination provides results that are closest to the all-atom calculations. We see that the $^3E$ -- $^3A_2$ is the most sensitive to the basis while the other two are less. Compared with the most localized case (\{16,16\}), the other results are consistently underestimated by up to 0.55 eV. From these results, it is clear that the extent of orbital localization affects various observables differently. While some optical transitions for a given system are insensitive, others can be extremely dependent on the basis. The sensible strategy is to construct the maximal fragment that also provides the maximal localization on each atom of interest and seek convergence of the observables of interest.

\begin{table}[H]
\centering
\caption{Excited-state transition energies of the NV$^-$ center in the 215-atom system using the Wannier function basis obtained from different sizes of the fragment as well as the all-atom calculation.}
\label{tab:table2}
\resizebox{\textwidth}{!}{\begin{tabular}{cccccc}
\toprule
\multirow{2}{*}{\begin{tabular}[c]{@{}c@{}}Transition\\ symmetry\end{tabular}} & \multicolumn{5}{c}{Energy (eV)}                               \\ \cline{2-6}
                                            & \{4,4\} & \{4,16\} & \{16,16\} &\{40,40\} &all-atom          \\ \midrule
$^3E$ -- $^3A_2$                            & 1.560 & 1.770 & 2.108 & 1.860 & 1.715  \\
$^1A_1$ -- $^3A_2$                          & 1.325 & 1.373 & 1.433 & 1.384 & 1.355 \\
$^1E$ -- $^3A_2$                            & 0.378 & 0.407 & 0.447 & 0.417 & 0.398 \\ 
$\sum_{i=1}^{4}L_i$                         & 3.514 & 3.464 & 3.507 & 3.411 & 3.461 \\  \bottomrule
\end{tabular}
}
\end{table}

In the rest of the paper, we employ the \{16,16\} fragment to obtain the PMW function basis and provide the detailed results in Tables~\ref{tab:table_215_16}$-$\ref{tab:table_2303_16}. The excited-state transition energies are summarized in Table~\ref{tab:table1}. For the bulk systems with the new ``p"-like basis, $^3E$ -- $^3A_2$ transition gap is enlarged by up to 0.6 eV from the less localized basis, while the other two transition energies are relatively less sensitive to the change of basis. These results again agree with various other theoretical calculations that employ PBE functionals to compute the bare Hubbard model parameters\cite{Goss1996,Gali2008,Delaney2010,Ma2010,Gordon2013,Alkauskas2014a}.

The effect of the fragment size is most pronounced for the slab. If the \{16,16\} fragment is used, the results are similar to those for the bulk systems. In detail: the $^3E$ -- $^3A_2$ transition is predicted $\sim$1 eV lower than that in the bulk, while the other two are only slightly lower (by $\sim 0.1$~eV) compared to the 999-atom cell. Here, the significant lowering of the triplet-triplet transition energy in the slab can be attributed to the interplay with the surface states of N atom passivation layer. The surface states dive below the conduction band minimum of the bulk states and are located inside the band-gap and affect the position of the in-gap defect states. Finally, we remark that these observations underline the importance of fragment selection. However, they are completely independent of the proposed sG-PMWF methodology. Indeed, the results obtained with the sG-PMWF and G-PMWF methods are perfect (Table~\ref{tab:ese_full}) in each case, while the results depend on the fragment size.

\section{Conclusions and Perspective}
By introducing the fragmentation treatment and the sequential exhausting of the orbital space to the traditional G-PMWF method, we develop a swift, efficient, and robust algorithm, sG-PMWF, to obtain a set of regionally localized states on a subsystem of interest. The completeness and efficiency are insensitive to the choice of input parameters. The core idea is to reduce the dimensionality of matrices during the maximization process and leads to a reduced scaling from being hyper-quadratic to linear. For the applications of localized basis to the Hubbard model, the excited-state calculations are sensitive to the localized basis. While Pipek-Mezey scheme is an ideal candidate to provide localized states with optimal localization for the whole system, it does not necessarily leads to ``maximally" localized orbitals on a specific subsystem. But in our fragmentation treatment, one can carefully select the atoms (the strategy is mentioned above) to reach ``maximally" localized orbitals on the subsystem as well as avoiding the over-localization issue. 

The resulting sG-PMWF method has five primary benefits: (1) largely shortens the time per SA iteration and makes it easier to monitor the progress of localization; (2) significantly lowers the total job time and scaling for systems with thousands of electrons;  (3) provides regionally localized orbitals with higher extent of localization; (4) less demanding for computing resources, e.g., memory and CPUs; (5) can be performed without the knowledge of canonical eigenstates if it is coupled with stochastic methods (e.g., stochastic DFT). The stochastic basis search approach exhibits higher efficiency than the traditional method for systems with over 4000 electrons.

Furthermore, we want to comment on the following prospective applications of the sequential exhausting method:
(1) this method can be generalized to obtain localized states of the whole system. Given that the rest space can always be updated or reconstructed by Gram-Schmidt orthogonalzaiton, the sG-PMWF calculation can then be sequentially applied to all the fragments in the whole system;
(2) this method can be coupled with other maximizer, e.g., conjugated gradient and BFGS approach, to further facilitate the convergence of the PM functional.

We believe that the sG-PMWF method will find numerous applications in condensed matter problems, either in chemistry, materials science, or computational materials physics.

\begin{acknowledgement}
This material is based upon work supported by the U.S. Department of Energy, Office of Science, Office of Advanced Scientific Computing Research, Scientific Discovery through Advanced Computing (SciDAC) program under Award Number DE-SC0022198. This research used resources of the National Energy Research
Scientific Computing Center, a DOE Office of Science User Facility
supported by the Office of Science of the U.S. Department of Energy
under Contract No. DE-AC02-05CH11231 using NERSC award
BES-ERCAP0020089.
\end{acknowledgement}

\begin{suppinfo}
The Supporting Information provides additional texts, figures and tables listed below.

Texts: Downfolded effective Hamiltonian for the Hubbard Model; Excited states of the NV$^-$ center; Preparation of stochastic basis using deterministic eigenstates.

Figures: three fragments of different sizes and the all-atom system exemplied by the 215-atom cell; scaling of $t^{outer}$ with respect to $N_w$, evolution of the objective functional with respect to the outer-loop step using deterministic and stochastic basis, investigations of ($N_c,N_r$) on the 511-atom system, total wall time of orbital localization on the four investigated systems, relative number of SA iterations in the sG-PMWF calculations for 215- and 511-atom systems, scaling of time per iteration with respect to $N_s$, electron density localized on the NV$^-$ center of the three solid systems, four ``p-like" localized Wannier functions on the NV$^-$ center of the three solid systems, electron density localized on an arbitrary carbon atom of the four investigated systems, four ``p-like" localized Wannier functions as well as the resulting electron density on the NV$^-$ center of the 215-atom system using different sizes of fragments; 

Tables: transition energies obatined from different Wannier functions basis using energy-windowing truncated space, comparison of G-PMWF and sG-PMWF calculations with different $N_c$ and $N_r$ for the four investigated systems using the 4-atom fragment, comparison of the time spent on the folding and unfolding steps; time per outer-loop iteration and normalized time per outer-loop iteration in the sG-PMWF calculations of the four investigated systems, total wall time and normalized total wall time of G-PMWF and sG-PMWF calculations for the four investigated systems, time per steep-ascent step in the G-PMWF and sG-PMWF calculations, number of iterations required to reach convergence for the four investigated systems, converged maximized PM functional values, spatial overlap between the Wannier functions obtained from the fragment approaches with those obtained from the all-atom calculation, transition energies of the four investigated systems using the Wannier functions basis obtained from the G-PMWF calculations, comparison of G-PMWF and sG-PMWF calculations with different $N_c$ and $N_r$ for the four investigated systems using the 16-atom fragment.
\end{suppinfo}

\bibliography{biblio}

\end{document}